\documentclass[12pt]{article}
\usepackage{amssymb,amsmath,slashed,latexsym}
\usepackage{float}
\allowdisplaybreaks
\newcommand{\ft}[2]{{\textstyle\frac{#1}{#2}}}

\def\rme{{\rm e}}
\def\rmi{{\rm i}}

\newcommand{\hc}{{\rm h.c.}}
\newcommand{\bbox}{\lower.2ex\hbox{$\Box$}}
\newsavebox{\uuunit}
\sbox{\uuunit}
    {\setlength{\unitlength}{0.825em}
     \begin{picture}(0.6,0.7)
        \thinlines
        \put(0,0){\line(1,0){0.5}}
        \put(0.15,0){\line(0,1){0.7}}
        \put(0.35,0){\line(0,1){0.8}}
       \multiput(0.3,0.8)(-0.04,-0.02){12}{\rule{0.5pt}{0.5pt}}
     \end {picture}}

\csname @addtoreset\endcsname{equation}{section}

\newcommand{\U}{\mathop{\rm {}U}}

   \usepackage[nosort]{cite}
   \pdfoutput=1
  \usepackage[pdftex]{hyperref}
\usepackage{tcolorbox}

\usepackage{color}
 
\newcommand{\poinc}{\boxdot}
\newcommand{\dr}{\raise.3ex\hbox{$\stackrel{\leftarrow}{\delta  }$}{}}
\newcommand{\dl}{\raise.3ex\hbox{$\stackrel{\rightarrow}{\delta }$}{} }
\newcommand{\pl}{\raise.3ex\hbox{$\stackrel{\rightarrow}{\partial }$}{} }
\newcommand{\chiomega}{\omega }
\newcommand{\pom}{{\rm om}}
\begin{document}

\begin{titlepage}
\begin{flushright}
CERN-TH-2018-071
\end{flushright}
\vspace{.5cm}
\begin{center}
\baselineskip=16pt
{\LARGE  Comments on rigid and local supercurrents\\[5mm] in ${\cal N}=1$ minimal Supergravity  
}\\
\vfill
{\large  {\bf Sergio Ferrara}$^{1}$, {\bf Marine Samsonyan}$^2$, \\[2mm] {\bf Magnus Tournoy}$^3$ and {\bf Antoine Van Proeyen$^3$}, } \\
\vfill

{\small$^1$ Theoretical Physics Department, CERN CH-1211 Geneva 23, Switzerland\\\smallskip
 INFN - Laboratori Nazionali di Frascati Via Enrico Fermi 40, I-00044 Frascati, Italy\\\smallskip
 Department of Physics and Astronomy and Mani L.Bhaumik Institute for Theoretical Physics, U.C.L.A., Los Angeles CA 90095-1547, USA\\\smallskip
 $^2$ CANDLE SRI, 31 Acharyan, 0040 Yerevan, Armenia \\\smallskip
$^3$   KU Leuven, Institute for Theoretical Physics, Celestijnenlaan 200D, B-3001 Leuven,
Belgium  \\[2mm] }
\end{center}
\vfill
\begin{center}
{\bf Abstract}
\end{center}
{\small
We discuss local supercurrents as sources of the super-Einstein equations in the superconformal approach in the old and new minimal (auxiliary fields) formulation. Modifications of the Ward identity giving the covariant divergence of the Einstein multiplet are considered in presence of a Fayet--Iliopoulos term. Curvature multiplets can be used as alternative to the gravitino variation in the search for rigid supersymmetric curved backgrounds.
} \vfill

\hrule width 3.cm
{\footnotesize \noindent e-mails: Sergio.Ferrara@cern.ch, Marine.Samsonyan@cern.ch, magnus.tournoy@kuleuven.be,\\
antoine.vanproeyen@fys.kuleuven.be }
\end{titlepage}

\addtocounter{page}{1}
 \tableofcontents{}
\newpage
\section{Introduction}

In this paper we consider the supercurrent multiplet when coupled to supergravity, using superconformal and superspace techniques. The results presented here are an extension of a previous investigation of the same authors \cite{Ferrara:2017yhz}. In rigid supersymmetry the supercurrent multiplet has a certain amount of ambiguity due to the possibility of improvement terms. These redefinitions modify the conservation law of a supercurrent $J_{\alpha \dot \alpha }$ expressed through its supercurrent divergence \cite{Ferrara:1975pz,Ferrara:1977mv,Gates:1983nr,Komargodski:2009pc,Komargodski:2010rb,Kuzenko:2010ni}
\begin{equation}
  \overline{D}^{\dot \alpha }J_{\alpha \dot \alpha }\approx  D_\alpha  Y +\chiomega _\alpha \,,
 \label{conservationEqn}
\end{equation}
where $Y$, $\chiomega _\alpha $ are chiral superfields and moreover $\chiomega _\alpha $ satisfies a reality condition $D_\alpha \chiomega ^\alpha = \overline{D}_{\dot \alpha }\chiomega ^{\dot \alpha}$.
In linearized supergravity \cite{Ferrara:1977mv,Komargodski:2010rb} the current couples to the Einstein multiplet $E_{\alpha \dot \alpha }$:\footnote{We will indicate equations valid modulo equations of motion by $\approx $.}
\begin{equation}
  E_{\alpha \dot \alpha }\approx -\kappa ^2 J_{\alpha \dot \alpha }\,,
 \label{E+J0}
\end{equation}
together with a Ward identity, which (without a gauge multiplet) is
\begin{align}
  \mbox{old minimal supergravity: }\qquad \overline{D}^{\dot \alpha }E_{\alpha \dot \alpha }= & D_\alpha {\cal R}\,, \nonumber\\
  \mbox{new minimal supergravity: }\qquad \overline{D}^{\dot \alpha }E_{\alpha \dot \alpha }= & W_\alpha ^L\,,
\label{WISugra}
\end{align}
where ${\cal R}$ is a scalar chiral multiplet and $W_\alpha ^L$ is a spinor chiral multiplet. The trace equation becomes in the two formulations
\begin{equation}
  {\cal R}\approx -\kappa ^2 Y\,,\qquad W_\alpha ^L\approx -\kappa ^2\chiomega  _\alpha\,.
 \label{XYW}
\end{equation}
The equations (\ref{E+J0}) and (\ref{XYW}) are the supergravity extensions of the Einstein equation and its trace
\begin{align}
  G_{\mu \nu }\approx &\ \kappa ^2 T_{\mu \nu }\,,\nonumber\\
  R \approx&  -\kappa ^2 T_\mu {}^\mu \,,\label{traceEinstein}
\end{align}
Equations (\ref{WISugra}) also contain the Einstein tensor identity (strong equation $=$):
\begin{equation}
  \nabla ^\mu G_{\mu \nu }=0\,,
 \label{nablaG0}
\end{equation}
which then implies the `weak' equation ($\approx $)
\begin{equation}
  \nabla ^\mu T_{\mu \nu }\approx 0\,,
 \label{nablaT0}
\end{equation}
The latter is the covariant conservation law of the stress tensor, which holds when the matter field equations are imposed.
Note that in the superconformal supergravity context the Planck constant is hidden in the value of the (first component of the) compensating multiplet, which is a chiral superfield in old minimal supergravity and a linear multiplet in new minimal supergravity
\begin{align}
  \mbox{old minimal supergravity: }&\qquad \overline{{\cal D}}_{\dot \alpha} X^0=0\,,  &X^0\stackrel{\boxdot}{=}\kappa ^{-1} \nonumber\\
  \mbox{new minimal supergravity: }&\qquad  T(L)=\bar{T}(L)=0\,,&L\stackrel{\boxdot}{=}\kappa ^{-2}\,,
\label{compensating}
\end{align}
where we denote by $\stackrel{\boxdot}{=}$ equations due to the (super)conformal gauge fixing. $T$ is the superconformal version of the superspace operator $\overline D^2$ \cite{Kugo:1983mv,Ferrara:2016een}, which defines a chiral multiplet from another multiplet, and $\overline{T}$ defines an antichiral multiplet.
\subsection{Old minimal Superconformal formulation: results}
It has been speculated in the literature that the gravity analogue of the conservation law (\ref{conservationEqn}) where both $Y$ and $\chiomega _\alpha$ are present corresponds to a new type of supercurrent with $(16+16)$ bosonic + fermionic degrees of freedom, while the two minimal supergravities (\ref{WISugra}) have $(12+12)$ degrees of freedom and the maximal non-minimal supergravity has $(20+20)$ degrees of freedom \cite{Gates:1983nr,Komargodski:2010rb,Kuzenko:2010ni}. In the present paper we will argue that the conservation law when both $Y$ and $\chiomega _\alpha $ are not vanishing is perfectly possible whenever an additional gauge symmetry acts on the chiral compensating multiplet $X^0$. This just happens in old minimal supergravity when a Fayet-- Iliopoulos (FI) term and a K\"{a}hler potential appear.

The basic Lagrangian we consider is
\begin{equation}
  {\cal L}= \left[N(X^I,\bar X^I) \rme^{\xi V}\right]_D + \left[W^\alpha W_\alpha \right]_F\,,\qquad \{X^I\}=\{X^0,\,X^i\}\,,
 \label{Loldmin}
\end{equation}
where $\xi $ is the (dimensionless) FI constant.
We will assume that the fields $X^i$ appear only in the form
\begin{equation}
  N(X^I,\bar X^I) = X^0\bar X^0\Phi (S^i,\bar S^i)\,,\qquad S^i \equiv X^i/X^0\,,
 \label{NXS}
\end{equation}
Note that under FI $\U(1)$ gauge transformations
\begin{equation}
  X^0\ \rightarrow\ X^0 \rme^{-\xi \Lambda }\,,\qquad V\ \rightarrow \ V + \Lambda +\bar \Lambda \,,\qquad S^i\ \mbox{ inert,}
 \label{LambdaWZ}
\end{equation}
while under K\"{a}hler transformations
\begin{equation}
  X^0\ \rightarrow\ X^0 \rme^{\Xi }\,,\qquad \Phi \ \rightarrow \ \Phi \rme^{-\Xi -\overline{\Xi }}\,, \qquad W(S)\ \rightarrow \ W(S)\rme^{-3\Xi}\,,
 \label{Kahlertransfo}
\end{equation}
where we mention the transformation of the superpotential for completeness. We do not further consider a superpotential in this paper.
However, since in this paper we will confine to an additive $\Phi $ such that
\begin{equation}
  \Phi = -3 +\Phi ^{\rm M}\,,
 \label{Phisplit}
\end{equation}
where the first term corresponds to pure supergravity, the K\"{a}hler invariance will be lost. Below, we will see that in the new minimal formulation,
unlike the old minimal one, the gravity-matter splitting preserves the K\"{a}hler invariance. See below (\ref{Snmmatter}), where the K\"{a}hler invariance is evident.

The super-Einstein equations are obtained from the Lagrangian (\ref{Loldmin}) by variation of the auxiliary field $A_\mu $ (the gauge field of the $\U(1)$ R-symmetry of the superconformal algebra). So the super-Einstein equation becomes\footnote{We follow the conventions of \cite{Freedman:2012zz}, where the actions, transformations and field equations are given explicitly. Some additional notation is introduced in \cite{Ferrara:2016een}.}
\begin{equation}
  \ft14\rmi \gamma ^\mu _{\alpha \dot \alpha } e^{-1}\frac{ \delta {\cal L}}{\delta A^\mu }=  {\cal E}^{\pom,V}_{\alpha \dot \alpha }(X^0,V) + J_{\alpha \dot \alpha }(X^0,V,S^i)+{\cal E}^W_{\alpha \dot \alpha }\approx 0\,,
 \label{Esplit}
\end{equation}
where we split the contributions according to (\ref{Loldmin}) and further to (\ref{Phisplit}), and we have
\begin{equation}
  {\cal E}^W_{\alpha \dot \alpha }=4W _{\dot \alpha }W_\alpha\,,\qquad
  W_\alpha \equiv  T {\cal D}_\alpha V= -\ft12\rmi \lambda _\alpha \,,
 \label{EWintro}
\end{equation}
$\lambda _\alpha $ being the left-projected gaugino in the gauge multiplet.

We will prove a `strong identity' ($V$, $X^0$ arbitrary)
\begin{equation}
 \overline{{\cal D}}^{\dot \alpha }{\cal E}^{\pom,V}_{\alpha \dot \alpha }
   =(\rme^{\xi V}X^0)^3{\cal D}_\alpha \left[\frac{\rme^{-3\xi V}T\left(\rme^{\xi V}\bar X^0\right)}{(X^0)^2}\right]+ 3 \xi X^0 \rme^{\xi V}\bar X^0 W_\alpha \,.
 \label{DEoldminFIVintro}
\end{equation}
The superspace geometry that encodes the Ward identity in (\ref{DEoldminFIVintro}) was denominated `chirally extended supergravity' in \cite{deWit:1978ww}.
Note that for $\xi =0$ this becomes the Ward identity of \cite{Ferrara:2017yhz}. Equation (\ref{DEoldminFIVintro}) has both terms as in (\ref{conservationEqn}) even if we are in old minimal supergravity.
The field equations coming from (\ref{Esplit}) have a contribution coming from the $V$ and $X^0$ field equations (field equations for the auxiliary fields, respectively $D$ and $ F^0$)
\begin{align}
 &\ft12D= {\cal D}_\alpha W^\alpha = \overline{{\cal D}}_{\dot \alpha} \bar W^{\dot \alpha}\approx -\ft14\xi  N\rme^{\xi V}=-\ft14\xi  (N^{{\rm{G}}} + N^{\rm{M}})\rme^{\xi V} \,,\nonumber\\
 & T(N_0\rme^{\xi V})\approx 0\,,
 \label{eomV}
\end{align}
where we have introduced
 \begin{equation}
  N= N^{{\rm{G}}} + N^{\rm{M}}\,,\qquad N^{\rm{G}} = -3 X^0\bar X^0\,,\qquad N^{\rm M}= X^0\bar X^0\Phi ^{\rm M}(S^i,\bar S^i)\,,
 \label{Nsplit}
\end{equation}
while
\begin{equation}
  N_0\equiv \frac{\partial N(X,\bar X)}{\partial X^0} = \bar X^0 \left(\Phi -S^i\frac{\partial \Phi }{\partial S^i}\right)= -3\bar X^0+N_0^{\rm M}\,,\qquad  N_0^{\rm M}= \bar X^0 \left(\Phi^{\rm M} -S^i\frac{\partial \Phi^{\rm M}}{\partial S^i}\right)\,.
 \label{N0}
\end{equation}
The second line in (\ref{eomV}) gives the trace of the Einstein equations in de Sitter space as described in section \ref{ss:FIcomponents}.
Acting with $\overline{{\cal D}}^{\dot \alpha }$ on  (\ref{EWintro}) using (\ref{eomV}) we find
\begin{equation}
     \overline{{\cal D}}^{\dot \alpha }{\cal E}^W_{\alpha \dot \alpha }=-2W_\alpha \,D  \approx \xi W_\alpha \,  (N^{{\rm{G}}} + N^{\rm{M}})\rme^{\xi V}\,.
 \label{DEWintro}
\end{equation}
From the field equation (\ref{Esplit}) we have
\begin{equation}
  \overline{{\cal D}}^{\dot \alpha }{\cal E}^{\pom,V}_{\alpha \dot \alpha } + \overline{{\cal D}}^{\dot \alpha }J_{\alpha \dot \alpha }+ \overline{{\cal D}}^{\dot \alpha }{\cal E}^W_{\alpha \dot \alpha }\approx 0\,.
 \label{sumDE}
\end{equation}
In this sum, the first term of (\ref{DEWintro}) (with $N^{{\rm{G}}}$) cancels the last term in (\ref{DEoldminFIVintro}) and with the second part of  (\ref{eomV}) we obtain
\begin{equation}
 \overline{{\cal D}}^{\dot \alpha }J_{\alpha \dot \alpha }
   \approx \ft13 (\rme^{\xi V}X^0)^3{\cal D}_\alpha \left[\frac{\rme^{-3\xi V}T\left(\rme^{\xi V}N_0^{\rm{M}}\right)}{(X^0)^2}\right]-\xi  W_\alpha N^{\rm{M}}\rme^{\xi V}\,,
 \label{DJmatterFI}
\end{equation}
which is the generalization of $R\approx -\kappa ^2T_\mu {}^\mu $ of general relativity. We understand the last term in the context of electromagnetism in Appendix \ref{ss:currentem}.

Note that for conformally coupled (neutral) chiral multiplets $N_0^{\rm{M}}=0$ and the violation of the superconformal symmetry comes only from the last terms due to the FI term.

\subsection{New minimal Superconformal formulation: results}
In the new minimal formulation with neutral matter and FI term, the action is
\begin{equation}
  {\cal L}^{\rm nm}= \left[ 3L\ln\frac{L}{X^0\bar X^0}\right]_D+ \left[L\,K\right]_D +\xi \left[L\, V\right]_D + \left[W^\alpha W_\alpha \right]_F\,,
 \label{Snmmatter}
\end{equation}
where $K$ is the matter K\"{a}hler potential
\begin{equation}
  K(S,\bar S) = -3\ln\left(-\ft13\Phi (S,\bar S)\right)\,.
 \label{KSbarS}
\end{equation}
The relevant field equation is here the field equation of $\chi _\alpha $, the fermionic component of the linear multiplet, which defines $W_\alpha ^{L}$, $W_\alpha ^{K}$ and $W_\alpha ^{\rm nm,FI}$ for the three first terms of (\ref{Snmmatter})\footnote{e.g. $W^L_\alpha$ is $\rmi$ times the covariant field equation for $P_L\chi $ of the first term of (\ref{Snmmatter}), see  Appendix \ref{app:linWL}.}
\begin{align}
 & W_\alpha ^{L}+W_\alpha ^{K}+W_\alpha ^{\rm nm,FI}\approx 0\,,\nonumber\\
 &W_\alpha^L=3\, T {\cal D}_\alpha \ln \frac{L}{X^0\bar X^0}\,,\nonumber\\
  &W_\alpha ^{K}= T {\cal D}_\alpha K= -\ft12\rmi\lambda_\alpha  (K)\,,\nonumber\\
  &  W_\alpha ^{\rm nm,FI}=\xi W_\alpha \,,
 \label{fechitotal}
\end{align}
where $\lambda _\alpha  (K)$ is the $P_L\lambda$ component of the real multiplet $K$.\footnote{The superfield $W_\alpha ^K$ is also the main ingredient in the construction of super-K\"{a}hler invariant actions in \cite{Cecotti:1986qw} and in the new actions for broken supergravity, dubbed `Liberated supergravity' in \cite{Farakos:2018sgq}. In these papers, this superfield is used in the context of old minimal supergravity.} Clearly, $\chi $ does not appear in the last term of (\ref{Snmmatter}).

The full $A_\mu $ field equation is split as
\begin{equation}
  {\cal E}^L_a+ J_a+{\cal E}^W_a\approx 0\,,
 \label{fullEnmK}
\end{equation}
where $J_a$ is related to the part with the K\"{a}hler potential in (\ref{Snmmatter}) (the FI term does not contribute).
We will prove below the strong equations
\begin{equation}
  \overline{{\cal D}}^{\dot \alpha }{\cal E}^L_{\alpha \dot \alpha }= L\,W^L_\alpha \,,\qquad  \overline{{\cal D}}^{\dot \alpha }{\cal E}^W_{\alpha \dot \alpha }=-2W_\alpha \,D \,.
 \label{DELintro}
\end{equation}
The $D$ field equation is now $2D+\xi L\approx 0$, leaving us with
\begin{equation}
  \overline{{\cal D}}^{\dot \alpha }J_{\alpha \dot \alpha }\approx  L\, W^K _\alpha  \,,
 \label{finalconsnmK}
\end{equation}
which is identical to the rigid equation for the supercurrent of the matter system.
\bigskip

In section \ref{ss:WI} we will give a general proof of the strong equations (Ward identities) mentioned above, based on general rules of the transformations of field equations.
We obtain the higher $\theta $ components of the Einstein multiplet in section \ref{ss:Einsteincomponents}, and in section \ref{ss:apply} we discuss applications related to rigid supersymmetric curved backgrounds. We give an account of the different auxiliary field formulations and related conformal compensators of ${\cal N}=1$ supergravity in section \ref{app:auxf}.
We finish with a conclusion in Section \ref{ss:conclusion}. In Appendix \ref{app:linWL}, we give some information on the linear multiplet and the related fermionic chiral superfield.
Superconformal Ward identities are given in Appendix~\ref{ss:WardId}. Appendix \ref{ss:currentem} explains an electromagnetic analogue of the conservation law. In Appendix \ref{app:chiralsgcoupling} we discuss the example where supergravity with FI term is coupled to a charged chiral multiplet, and we consider also a nilpotent chiral multiplet so that the model includes a Volkov--Akulov field.
\section{Ward identities}
\label{ss:WI}
\subsection{General equation}
\label{ss:genConsEqn}
As mentioned in the introduction, we obtain results by giving first Ward identities that are of the form
\begin{equation}
  \overline{{\cal D}}^{\dot \alpha }{\cal E}_{\alpha \dot \alpha }= \ldots\,,\qquad {\cal E}_{\alpha \dot \alpha }=\ft14\rmi \gamma ^a _{\alpha \dot \alpha }{\cal E}_a \,,\mbox{ or }{\cal E}_a  = 2\rmi {\cal E}_{\alpha \dot \alpha }(\gamma _a)^{\alpha \dot \alpha}\,,
 \label{conservationEqnmodified}
\end{equation}
where ${\cal E}_{\alpha \dot \alpha }$ or, equivalently
${\cal E}_a $, is defined by the field equation of $A_\mu$ in supergravity actions: in general
\begin{equation}
{\cal E}_a \equiv  -\frac43 e^{-1}e_a^\mu  \frac{\delta }{\delta A^\mu }S   \,.
 \label{calEdefined}
\end{equation}
In the notation where $\Theta (\phi ^i)$ indicates the covariantized field equation\footnote{$\Theta (\phi) _i\equiv \frac{\dl S}{\delta \phi ^i}+$ non-covariant terms.} of the field $\phi ^i$, the latter equation is written
\begin{equation}
  {\cal E}_a  = -\frac{4}{3}\Theta (A_a)\,.
 \label{defEmu}
\end{equation}
The multiplet ${\cal E}_a $ has (Weyl,chiral) weights (3,0). The right-hand side of (\ref{conservationEqnmodified}) should have weights $(7/2,3/2)$. Other components of ${\cal E}_a $ are obtained by successive supersymmetry transformations on this lowest component of the multiplet.

We will consider several actions, e.g. $S^{\pom}$ and define then $ {\cal E}_a^{\pom}$ as the expression defined by (\ref{calEdefined}) when $S=S^{\pom}$.
But first we find general relations that do not depend on the choice of action $S$.

The quantity in (\ref{conservationEqnmodified}) is
\begin{equation}
  \overline{{\cal D}}^{\dot \alpha }{\cal E}_{\alpha \dot \alpha }=  -\ft14\rmi\left(\gamma ^a\delta _R{\cal E}_a\right)_\alpha \,,
 \label{defDalEaldotal}
\end{equation}
where $\delta _R$ is defined by
\begin{equation}
   \delta {\cal E}_a = \bar \epsilon P_L \delta _L {\cal E}_a+ \bar \epsilon P_R \delta _R {\cal E}_a\,.
 \label{defdeltaLR}
\end{equation}
To find this, we use a method explained in \cite[Appendix D]{Ferrara:2017yhz}, based upon earlier work in e.g.  \cite{VanProeyen:1983wk,Vanhecke:2017chr}.
The covariant field equations satisfy Ward identities
\begin{equation}
  \delta(\epsilon )\phi ^j\,\Theta(\phi )_j =0\,.
 \label{WIgeneral}
\end{equation}
We will use the formula that gives the symmetry transformation of covariant field equations
\begin{equation}
  \delta (\epsilon )\Theta(\phi ^i) = -\left(\frac{\pl \delta(\epsilon )\phi ^j}{ \partial \phi ^i}\right)_{\rm cov}\,\Theta(\phi )_j\,,
 \label{delThetacov}
\end{equation}
where `cov' refers to a covariantization of all spacetime derivatives and omission of terms with undifferentiated gauge fields.
The equation depends on which multiplets are taken in the theory and on their transformation laws (see the sum over $j$ in the right-hand side).
E.g. since $A_\mu $ does not appear in $S$-supersymmetry transformations of other fields,  (\ref{delThetacov}) with $\epsilon $ replaced by the $S$-supersymmetry parameter $\eta $ implies that  (\ref{defEmu}) does not transform under $S$-supersymmetry and therefore $\Theta (A_a)$ defines a superconformal primary.

We will consider this now for superconformal $Q$-supersymmetry, with\footnote{The generalization to several real or linear multiplets is obvious, adding indices that are summed similar to the sum over $I$ for the chiral multiplets.}
\begin{description}
  \item[the Weyl multiplet:] $\{e_\mu ^a,\,\psi _\mu  ,\, b_\mu ,\,A_\mu \}$\,.
  \item[chiral multiplets:] $\{X^I,\, \Omega ^I,\, F^I\}$.
  \item[a real vector multiplet:]
$\{V,\,\zeta ,\,{\cal H},\,W_a,\,\lambda ,\,D\}$, which in Wess--Zumino (WZ) gauge is reduced to $\{W_\mu ,\,\lambda ,\,D\}$.
  \item[a real linear multiplet:] $\{L,\,a_{\mu \nu },\,\chi\}$\,.
\end{description}

The supersymmetry transformation of the field equation of $A_\mu $ according to (\ref{delThetacov}), with $\phi ^i$ replaced by $A_\mu $, depends on the fields that have a term $A_\mu \epsilon $ in their transformation. Such terms are included in covariant derivatives. Considering all the fields mentioned above, we find
\begin{align}
  \delta (\epsilon )\Theta (A_a)=\bar \epsilon P_R&\left[-\frac32 \rmi\Theta (\psi )_a-\frac{1}{\sqrt{2}}\rmi\gamma _a\left( X^I\Theta (\Omega )_I+\frac12\Omega ^I\Theta (F)_I\right)\right.\nonumber\\
  &\left.-\frac32 \gamma _a\zeta \Theta ({\cal H})-\ft34\rmi\zeta \Theta (W)_a+\frac34\gamma _a\lambda \Theta (D)\right]+\hc\,.
 \label{delThetaA}
\end{align}
Combining this with (\ref{defDalEaldotal}) and (\ref{defEmu}), we find
\begin{align}
  \overline{{\cal D}}^{\dot \alpha }{\cal E}_{\alpha \dot \alpha }=&\left[\ft12 \gamma ^a\Theta (\psi )_a +\ft23\sqrt{2} X^I\Theta (\Omega )_I+\ft13\sqrt{2} \Omega ^I\Theta (F)_I\right.\nonumber\\
&\left. -2\rmi \zeta\Theta ({\cal H})+\ft14 \gamma ^a\zeta\Theta (W)_a +\rmi \lambda \Theta (D)
  \right]_\alpha \,.
 \label{generalDE}
\end{align}
This expression can be rewritten upon use of the Ward identity (\ref{WIgeneral}) for $S$-supersymmetry, see (\ref{Ward S-sym}),
\begin{equation}
 W(S)\equiv  (\delta _S\phi ^i)\, \Theta (\phi )_i = 0\,.
 \label{WSgeneral}
\end{equation}
Considering the mentioned multiplets, the left projection of this equation is
\begin{equation}
P_L  W(S) = P_L\left[\gamma ^a \Theta (\psi )_a +\sqrt{2}X^I\,\Theta (\Omega )_I-2\rmi \zeta \Theta ({\cal H})+\ft12\gamma ^a\zeta \Theta (W)_a-2\rmi L\,\Theta (\chi )\right]\,.
 \label{WS}
\end{equation}
Combining (\ref{generalDE}) and (\ref{WS}), we get
\begin{equation}
 \overline{{\cal D}}^{\dot \alpha }{\cal E}_{\alpha \dot \alpha }=\left[\ft16\sqrt{2}\left( X^I\Theta (\Omega )_I+2\Omega ^I\Theta (F)_I\right)-\rmi\zeta  \Theta ({\cal H})+\rmi \lambda \Theta (D)+\rmi L\,\Theta (\chi )\right]_\alpha \,.
 \label{generalDEafterS0}
\end{equation}
In Appendix \ref{app:linWL} it is shown that the general equation  (\ref{delThetacov}) also implies that
\begin{equation}
  W_\alpha ^{\chi }= \rmi \Theta (\chi )_\alpha\,,\qquad W_\alpha = -\ft12\rmi\lambda _\alpha\,,
 \label{defW2}
\end{equation}
satisfy the same properties (chiral primaries with real ${\cal D}^\alpha W_\alpha $). Therefore, we still rewrite this equation as
\begin{tcolorbox}
\begin{equation}
 \overline{{\cal D}}^{\dot \alpha }{\cal E}_{\alpha \dot \alpha }=\ft16\sqrt{2}\left[ X^I\Theta (\Omega )_{I\alpha }+2\Omega ^I_\alpha \Theta (F)_I\right]-\rmi\zeta_\alpha  \Theta ({\cal H})-2 W_\alpha  \Theta (D)+W^{\chi}_\alpha \,L\,.
 \label{generalDEafterS}
\end{equation}
\end{tcolorbox}
The upper indication $\chi $ for $W^{\chi}_\alpha $ will later be used to refer to the $\chi $ field equation of particular actions, e.g. $S^L$ for the pure supergravity new minimal model:  (\ref{SL}).
The term with $\zeta_\alpha   \Theta ({\cal H})$ is absent in WZ gauge.
The last two terms of (\ref{generalDEafterS}) are of the form of the second term in (\ref{conservationEqn}) when $\Theta (D)_A$ is a constant (FI constant) or $L$ is a constant (conformal gauge choice $L\stackrel{\boxdot}{=}\kappa ^{-2}$).

\subsection{Old minimal pure supergravity}
\label{ss:pom}
We first consider pure supergravity with only the compensating multiplet $X^0$. Thus we only have the terms with $\Theta (\Omega )_0$ and $\Theta (F)_0$ in the right-hand side of (\ref{generalDEafterS}).

When we apply (\ref{delThetacov}) for $\phi ^i= F^I$ we get
\begin{equation}
  \delta(\epsilon )\Theta ( F)_I = -\frac{1}{\sqrt{2}}\bar \epsilon P_L\Theta (\Omega )_I\,.
 \label{delepsThetaF}
\end{equation}
Therefore we have
\begin{equation}
  \delta (\epsilon ) \frac{\Theta ( F)_0 }{(X^0)^2} = -\frac{1}{\sqrt{2}(X^0)^3} \left[X^0\Theta (\Omega )_0 +2 \Omega ^0\Theta (F)_0\right]\,,
 \label{delepsthetaX0}
\end{equation}
The quantity in the left-hand side has (Weyl,chiral) weights~(0,0). Applying  (\ref{delThetacov}) for S-supersymmetry for $\phi ^i=F^I$ shows that $\Theta ( F)_I$ is S-supersymmetry invariant since $F^I$ does not appear in the S-transformations of the fields. Therefore ${\cal D}_\alpha $ is well defined on this quantity and we can write (\ref{generalDEafterS}) as
\begin{equation}
 \overline{{\cal D}}^{\dot \alpha }{\cal E}^{\pom}_{\alpha \dot \alpha }=-\frac13 (X^0)^3{\cal D}_\alpha \frac{\Theta ( F)_0 }{(X^0)^2}\,,
 \label{DEoldmin}
\end{equation}
where `om' stands for pure `old minimal' supergravity. We can apply it to the action
\begin{equation}
  S^{\pom}= \left[-3 X^0\bar X^0\right]_D\,,
 \label{Spom}
\end{equation}
where $\Theta ( F)_0= -3\bar F^0= -3 T(\bar X^0)$. With the definition
\begin{equation}
  {\cal R}\equiv \frac{1}{X^0} T(\bar X^0)\,,
 \label{calRdefined}
\end{equation}
we thus get the equation in the form presented in \cite{Ferrara:2017yhz}:
\begin{tcolorbox}
\begin{equation}
  \overline{{\cal D}}^{\dot \alpha }{\cal E}^{\pom}_{\alpha \dot \alpha }= (X^0)^3{\cal D}_\alpha \frac{{\cal R} }{X^0}\,.
 \label{DEpom}
\end{equation}
\end{tcolorbox}
The value of ${\cal E}^{\pom}_a$ is in this case
\begin{equation}
  {\cal E}^{\pom}_a= -8X^0\bar X^0 A_a +4\rmi X^0\partial _a\bar X^{\bar 0} -4\rmi \bar X^{\bar 0}\partial _a X^0 +2\rmi \bar \Omega^0P_L\gamma _a\Omega ^{\bar 0}
 + 2\sqrt{2}\rmi \bar \psi _a\left(P_L\Omega ^0X^{\bar 0}-P_R\Omega ^{\bar 0}X^0\right)
  \,,
 \label{valEpom}
\end{equation}
whose flat limit is
\begin{equation}
  \left.{\cal E}_{\alpha \dot\alpha  }^{\pom} \right|_{\rm flat}=
  - 4 \rmi \bar X^{\bar 0}{\partial}_{\alpha \dot\alpha }X^0
  +4 \rmi X^{\bar 0}{\partial}_{\alpha \dot\alpha }\bar X^0
  -2(D_\alpha X^0)(\overline D_{\dot\alpha  }\bar X^{\bar 0})\,.
 \label{Ealdotalflat}
\end{equation}

\subsection{Old minimum supergravity with a FI term}
\label{ss:omFI}
We now apply the above for the case when $X^0$ transforms under an Abelian gauge group gauged by a real multiplet $V$. The supergauge transformations are given in (\ref{LambdaWZ}),
where the chiral superfield $\Lambda $ reduces in the WZ gauge to $\Lambda = -\ft12\rmi\theta $ plus contributions to the supersymmetry (see e.g. (2.28) in \cite{Cremmer:1982en}).
The $\theta $ transformation is then a phase transformation on the compensating scalar with gauge field  $W_\mu $.
\begin{equation}
  \delta(\theta ) X^0 = \ft12\rmi \xi \theta X^0\,,\qquad \delta W_\mu =\partial _\mu \theta \,.
 \label{deltheta}
\end{equation}
The action that provides the FI term is the covariantization of (\ref{Spom}), and is indicated by `om,V':
\begin{equation}
S^{\pom,V}= \left[-3 X^0\rme^{\xi V}\bar X^0\right]_D\,.
 \label{SpomFI}
\end{equation}
This action leads to covariant field equations
\begin{equation}
  \Theta (F^0)^V = -3 T(\rme^{\xi V}\bar X^0)\,,\qquad \Theta ({\cal H})^V= \ft32\xi X^0 T (\rme^{\xi V} \bar X^0)\,,
 \label{ThetaV}
\end{equation}
in agreement with the Ward identity for the transformation with parameter the upper complex component of $\Lambda $:
\begin{equation}
  \ft12\Theta ({\cal H})-\xi \,X^0\,\Theta (F_0)=0\,.
 \label{WILambdaF}
\end{equation}
Since we still have only one chiral multiplet, the rewriting of the first two terms of (\ref{generalDEafterS}) in (\ref{DEoldmin}) is still valid.
The field equation for the auxiliary field $D$ of the gauge multiplet in WZ gauge gives the moment map, ${\cal P}$, and the supergauge-invariant form is
\begin{equation}
  \Theta (D)^V = -{\cal P}\,,\qquad  {\cal P}= \ft32\xi X^0 \rme^{\xi V}\bar X^0\,.
 \label{DcalP}
\end{equation}
Therefore we get
\begin{equation}
 \overline{{\cal D}}^{\dot \alpha }{\cal E}^{\pom,V}_{\alpha \dot \alpha }
   =(X^0)^3{\cal D}_\alpha \left[\frac{T\left(\rme^{\xi V}\bar X^0\right)}{(X^0)^2}\right]+ 2W_\alpha {\cal P}-\ft32\rmi\xi \zeta _\alpha X^0 T (\rme^{\xi V} \bar X^0)\,.
 \label{DEoldminFIV0}
\end{equation}
Since $\rme^{3\xi V}{\cal D}_\alpha \rme^{-3\xi V}= -\ft32\rmi\xi \zeta _\alpha$, the last term can be absorbed in the first one 
\begin{tcolorbox}
\begin{equation}
 \overline{{\cal D}}^{\dot \alpha }{\cal E}^{\pom,V}_{\alpha \dot \alpha }
   =(\rme^{\xi V}X^0)^3{\cal D}_\alpha \left[\frac{\rme^{-3\xi V}T\left(\rme^{\xi V}\bar X^0\right)}{(X^0)^2}\right]+ 2W_\alpha {\cal P}\,.
 \label{DEoldminFIV}
\end{equation}
\end{tcolorbox}

When using this in WZ gauge, the exponentials of $V$ do not appear, and the equation simplifies to (we denote the ${\cal E}_a$ field now as ${\cal E}_a^{\pom,\rm FI}$)
\begin{equation}
 \overline{{\cal D}}^{\dot \alpha }{\cal E}^{\pom,\rm FI}_{\alpha \dot \alpha }
 = (X^0)^3{\cal D}_\alpha \frac{{\cal R} }{X^0}+2 W_\alpha {\cal P}\,.
 \label{DEoldminFI}
\end{equation}
But one should take into account that the supersymmetry transformations of a chiral multiplet are modified, i.e.
\begin{equation}
  \delta(\epsilon) \bar F^0 = \ldots -\ft12\rmi\bar \epsilon\xi P_L\lambda \bar X^0\,,
 \label{delextraF0}
\end{equation}
and covariant spacetime derivatives get a $W_\mu $ connection. Thus the meaning of supercovariant derivatives ${\cal D}_\alpha$ is different, and  (\ref{delextraF0}) induces an extra term
\begin{equation}
  (X^0)^3{\cal D}_\alpha \frac{{\cal R} }{X^0}=(X^0)^3{\cal D}_\alpha \frac{\bar F^0 }{(X^0)^2} = \ldots -\ft12 \rmi\xi X^0\bar X^0P_L\lambda =\ldots  + \ft23 {\cal P}W_\alpha  \,.
 \label{extraDalphaF}
\end{equation}
The latter replaces the contribution in (\ref{DEoldminFIV}) where the derivatives ${\cal D}_\alpha $ and $T$ act on $V$.
\bigskip

We can write a supergauge-invariant version of  (\ref{Ealdotalflat})
\begin{align}
  \left.{\cal E}^{\pom,V}_{\alpha \dot\alpha  } \right|_{\rm flat}= &
  2 \rme^{\xi V} X^0 D_\alpha \rme^{-\xi V} \overline D_{\dot \alpha }\rme^{\xi V} \bar X^0
  -2\rme^{\xi V} \bar X^0\overline D_{\dot \alpha }\rme^{-\xi V} D_\alpha \rme^{\xi V} X^0
\nonumber\\
&-2\rme^{-\xi V} D_\alpha (\rme^{\xi V} X^0) \,\overline D_{\dot \alpha }(\rme^{\xi V} \bar X^0)\,.
 \label{EVflat}
\end{align}
This expression makes no sense in conformal language since e.g. ${\cal D}_\alpha $ can only be applied on a multiplet with $c=-w$. Thus this expression is only applicable to the rigid supersymmetry limit. Therefore we write $D_\alpha $ and not ${\cal D}_\alpha $, ...
But we can check the coefficient of the last term in (\ref{DEoldminFIV}) by using this flat limit. Indeed, $\overline{D}^{\dot \alpha }$ on (\ref{EVflat}) gives  a term\footnote{The conventions in Appendix~A of \cite{Ferrara:2017yhz} imply $T=\bar D^2 = -\bar D^{\dot \alpha }\bar D_{\dot \alpha }$ and we neglect all terms that have less than 3 derivatives on $V$.}
\begin{equation}
 \left. \overline{D}^{\dot \alpha }{\cal E}^{\pom,V}_{\alpha \dot\alpha  } \right|_{\rm flat}=\ldots +\ft83 W_\alpha {\cal P}\,.
 \label{flatDEV}
\end{equation}
This contains the contributions of the last term of (\ref{DEoldminFIV}) and of (\ref{extraDalphaF}).

In WZ gauge, the field equation of $A_\mu $ is the covariantized version of (\ref{valEpom}), which implies that there is an extra term:
\begin{align}
  {\cal E}^{\pom,\rm FI}_a=&-8X^0\bar X^0 A_a +4\rmi X^0 \hat{\partial } _a\bar X^0 -4\rmi \bar X^0\hat{\partial} _a X^0 +2\rmi \bar \Omega^0P_L\gamma _a\Omega ^{\bar 0}\nonumber\\
  &+ 2\sqrt{2}\rmi \bar \psi _a\left(P_L\Omega ^0X^{\bar 0}-P_R\Omega ^{\bar 0}X^0\right)\nonumber\\
  =& {\cal E}^{\pom}_a-\ft83W_a{\cal P}\,,
 \label{valEpomFI}
\end{align}
where $\hat{\partial} _a X^0= \partial _a X^0 -\ft12\rmi\xi  W_a X^0$. The transformation of ${\cal E}^{\pom,\rm FI}_a$ contains then the transformation of $W_a$
\begin{equation}
  \delta_{\rm extra}(\epsilon ){\cal E}^{\pom,\rm FI}_a= -\ft83{\cal P} \delta(\epsilon )W_a=  \ft43{\cal P}\bar \epsilon \gamma _a\lambda \,,
 \label{delextraE}
\end{equation}
which leads to the same term (\ref{flatDEV}).

\subsection{Gauge kinetic terms}
\label{ss:gaugekin}
Now we consider
\begin{equation}
 S^W= -\ft14\left[\bar \lambda P_L\lambda \right]_F=\left[W^\alpha W_\alpha \right]_F\,,
 \label{SW}
\end{equation}
which added to  (\ref{SpomFI}) gives the off-shell version of the Freedman model \cite{Freedman:1976uk,deWit:1978ww}.
The $A_\mu $ field equation of this action leads to
\begin{equation}
   {\cal E}^{W}_a= \rmi \bar \lambda \gamma _*\gamma _a\lambda \,,\qquad {\cal E}^W_{\alpha \dot \alpha }= \lambda _{\dot \alpha }\lambda _\alpha =4W _{\dot \alpha }W_\alpha\,.
 \label{EWa}
\end{equation}
Here there are no chiral multiplets, and the only contribution in the right-hand side of (\ref{generalDEafterS}) comes from
\begin{equation}
  \Theta^W (D) = D=\rmi {\cal D}^\alpha \lambda _\alpha = -2 {\cal D}^\alpha W_\alpha \,.
 \label{ThetaWD}
\end{equation}
Hence the equation is
\begin{tcolorbox}
\begin{equation}
   \overline{{\cal D}}^{\dot \alpha }{\cal E}^W_{\alpha \dot \alpha }=-2W_\alpha \,D  \,,
 \label{DEW}
\end{equation}
\end{tcolorbox}
\noindent which can easily be checked from the supersymmetry transformation of $\lambda $.
For the full model with FI term and kinetic terms, we thus get
\begin{tcolorbox}
\begin{equation}
 \overline{{\cal D}}^{\dot \alpha }\left({\cal E}^{\pom,V}+{\cal E}^W\right)_{\alpha \dot \alpha }
   =(\rme^{\xi V}X^0)^3{\cal D}_\alpha \left[\frac{\rme^{-3\xi V}T\left(\rme^{\xi V}\bar X^0\right)}{(X^0)^2}\right]+ 2W_\alpha ({\cal P}- D)\,.
 \label{DEVcomplete}
\end{equation}
\end{tcolorbox}
\noindent where the last factor is the $D$-field equation.

\subsection{New minimal pure supergravity}
\label{ss:nmp}
Here we have only a real linear multiplet $L$ (and the chiral multiplet $X^0$, which, however, does not appear in the action):
\begin{align}
  S^L =& \left[3L\ln\frac{L}{X^0\bar X^0}\right]_D\nonumber\\
  =&\frac32\int\mathrm{d}^4 x\, \rme \,\Big[-\Box^c L+\frac1{2L}\left(v_av^a+{\cal D}_a L{\cal D}^a L+\bar{\chi}\slashed{\cal D}\chi\right)+\frac1{4L^3}\bar{\chi}P_L\chi\bar{\chi}P_R\chi\nonumber\\
&\phantom{\frac32\int\mathrm{d}^4 x\, \rme \,[}+\frac12\rmi \bar{\psi}_\mu\gamma^\mu\gamma_*\left(\slashed{\cal D}\chi+\frac1{2L}\left(\rmi\gamma_*\slashed{v}-\slashed{\cal D}L\right)\chi+\frac1{2L^2}\chi\bar{\chi}\chi\right)\Big]\nonumber\\
&\phantom{\frac32\int\mathrm{d}^4 x\, \rme \,[}+\varepsilon^{\mu\nu\rho\sigma}(2A_\mu-\frac1L v_\mu-\frac1{4L^2}\rmi\bar{\chi}\gamma_*\gamma_\mu\chi+\frac1{2L}\bar{\psi}_\mu\chi)\partial_\nu a_{\rho\sigma}\,,
 \label{SL}
\end{align}
using fields and notations as in Appendix \ref{app:linWL}.
The $A_\mu$  field equation gives now
\begin{equation}
  {\cal E}^L_a = -4 v_a+\frac3{2L}\bar \chi\gamma _* \gamma _a\chi \,.
 \label{EL}
\end{equation}
According to (\ref{generalDEafterS}), this satisfies
\begin{tcolorbox}
\begin{equation}
  \overline{{\cal D}}^{\dot \alpha }{\cal E}^L_{\alpha \dot \alpha }= L\,W^L_\alpha \,,
 \label{DEL}
\end{equation}
\end{tcolorbox}
\noindent where now we use for $\Theta (\chi )$ the field equation for the action (\ref{SL}):
\begin{equation}
W^L_\alpha= \rmi\Theta(\chi)^L\,,\qquad \Theta(\chi)^L=\frac3{2L} P_L\slashed{\cal D}\chi+\frac3{4L^3}P_L\chi\bar\chi P_R\chi-\frac{3}{4L^2}\rmi\slashed{v}P_R\chi \,. \label{chifeq}
\end{equation}
In Poincar\'{e} gauge, $L=\kappa ^{-2}$ and $\chi =0$, this is
\begin{equation}
W^L_\alpha\stackrel{\boxdot}{=}-\frac1{2}P_L\gamma\cdot\widehat{R}'(Q)\,,\qquad \widehat{R}'_{\mu\nu}(Q)=2\left(\partial_{[\mu}-\frac32\rmi A_{[\mu}\gamma_*+\frac14\omega_{[\mu}^{ab}\gamma_{ab}+\frac14\rmi\kappa^2\gamma_*\gamma_{[\mu}\slashed{v}\right)\psi_{\nu]}\,,
\end{equation}
which agrees with what is found in \cite{Ferrara:1988pd,Ferrara:1988qxa}.

\subsection{The FI term in new minimal supergravity}
\label{ss:FItermnm}
The FI term in new minimal supergravity is
\begin{equation}
  S^{\rm nm, FI}= \xi \left[L\, V\right]_D.
 \label{SFI}
\end{equation}
$A_\mu $ does not appear in this action. Hence (\ref{generalDEafterS}) reduces here to
\begin{equation}
  0= -2W_\alpha  \Theta^{\rm nm, FI} (D)+W_\alpha ^{\rm nm,FI}\, L\,.
 \label{0DEL}
\end{equation}
Indeed, $\Theta^{\rm nm, FI} (D)= \ft12 \xi L$ and $W_\alpha ^{\rm nm,FI}=-\xi \,W_\alpha$.
The FI term in new minimal does therefore not contribute to ${\cal E}_a$ and this is consistent with the conservation equations.

We can consider then the total action
\begin{equation}
S^{\rm nm,total}=  S^L + S^{\rm nm, FI} + S^W\,.
 \label{Stotalnm}
\end{equation}
The conservation equation is then
\begin{equation}
    \overline{{\cal D}}^{\dot \alpha }\left({\cal E}^L+{\cal E}^W\right)_{\alpha \dot \alpha }= L\,W^L _\alpha -2 D\, W_\alpha  \,,
 \label{totalconservnm}
\end{equation}
Adding the zero result (\ref{0DEL}) and using  (\ref{ThetaWD}) this is
\begin{align}
    &\overline{{\cal D}}^{\dot \alpha }\left({\cal E}^L+{\cal E}^W\right)_{\alpha \dot \alpha }= L\,W^{\rm nm,total}_\alpha -2  W_\alpha \Theta^{\rm nm,total} (D) \,,\nonumber\\
    & W^{\rm nm,total}_\alpha= 
    W^L _\alpha+\xi \,W_\alpha\,,\qquad
   \Theta^{\rm nm,total} (D) = D+ \ft12\xi L\,,
 \label{totalconservnm0}
\end{align}
and thus the right-hand side also vanishes by the field equations.

\section{Components of the Einstein curvature multiplet}
\label{ss:Einsteincomponents}

The Einstein current multiplet is defined by the field equation of $A_\mu $, the $\U(1)$ gauge field in the Weyl multiplet, see (\ref{calEdefined}) or (\ref{defEmu}). This is the first component of a real multiplet since it is a superconformal primary (it does not transform under $S$-supersymmetry). Then all other components are defined by the transformations of this first component.
These components are denoted as
\begin{equation}
  \left\{{\cal C}_a,\, {\cal Z}_a,\, {\cal H}_a,\,{\cal B}_{ba},\,\Lambda _a,\, D_a\right\}\,,
 \label{componentsC}
\end{equation}
where the fermions ${\cal Z}_a$ and $\Lambda _a$ are Majorana spinors, while from the bosons only ${\cal H}_a$ is complex and the others are real.

We omit now the overall coefficient in (\ref{defEmu}), and start from ${\cal C}_a=\Theta (A)_a$. We then use (\ref{delThetacov}) to derive the further components. E.g. (\ref{delThetaA}) determines the ${\cal Z}_a$ component. Sometimes we use Ward identities (\ref{WIgeneral}) for several symmetries to rewrite expressions. In (6.3) of \cite{Ferrara:2017yhz} we obtained this result for the coupling of the Weyl multiplet to chiral multiplets. Now we will add a real multiplet and a linear multiplet as in section \ref{ss:genConsEqn}.

The relevant components are then
\begin{align}
{\cal C}_a=&\Theta(A)_a\,,\nonumber\\
{\cal Z}_a=&3\Theta(\psi)_a-\gamma_a\gamma\cdot\Theta(\psi)+\frac1{\sqrt2}\gamma_a\left(\Omega^I\Theta(F)_I+\hc\right)
-\frac32\rmi\gamma_*\gamma_a\lambda^A\Theta(D)_A-2\rmi\gamma_*\gamma_a\Theta(\chi)L\,,\nonumber\\
{\cal H}_a=&2\rmi \bar X^{\bar I}{\cal D}_a\Theta(F)_{\bar I}-4\rmi\Theta(F)_{\bar I}{\cal D}_a\bar X^{\bar I} +3\rmi\bar\lambda^A\gamma_a P_L\Theta(\lambda)_A+ 3\rmi\bar\chi\gamma_a P_R\Theta(\chi)\,,\nonumber\\
{\cal B}_{ba}=&3\Theta(e)_{ab}-\eta_{ab}\Theta(e)_c{}^c+\ft12\varepsilon_{abcd}{\cal D}^c\Theta(A)^d-\ft12\eta_{ab}\left(\bar\Omega^I\Theta(\Omega)_I+2F^I\Theta(F)_I+\hc\right)\nonumber\\
& +\eta_{ab}D^A\Theta(D)_A +3\rmi\tilde F_{ab}{}^A\Theta(D)_A +\ft32\bar\lambda^A\gamma_{ab}\Theta(\lambda)_A\nonumber\\
&-3L\Theta(a)_{ba}-2\eta_{ab}L\Theta(L)
-\bar\chi\Theta(\chi)-\ft32\bar\chi\gamma_{ab}\Theta(\chi)\,,\nonumber\\
\Lambda_{a}=&\;2\gamma^b {\cal D}_{[a} {\cal Z}_{b]}-3\sqrt{2}\left(\Theta(\Omega)^I {\cal D}_a X_I +\Omega^I {\cal D}_a \Theta(F)_I + \hc\right)\nonumber\\
&+\ft32\left(\ft12\gamma\cdot F^A-\rmi\gamma_* D^A\right)\gamma_a\Theta(\lambda)_A+\ft32\gamma_b\gamma_a\lambda^A\Theta(W)_A^b\nonumber\\
&+\ft32\rmi\gamma_*\gamma^b\gamma_a\lambda^A{\cal D}_b\Theta(D)_A -6\rmi\gamma _*{\cal D}_a\left(\lambda^A\Theta(D)_A\right)\nonumber\\
&-\ft32\left(\slashed{v}-\rmi\gamma_*\slashed{\cal D}L\right)\gamma_a\Theta(\chi)-\ft32\rmi\gamma_*\gamma_a\chi\Theta(L)-\ft34\rmi\gamma_*\gamma_{bc}\gamma_a\chi\Theta(a)^{bc}-6\rmi\gamma_*{\cal D}_a\left(\Theta(\chi)L\right)\,,\nonumber\\
D_{a}=&-2{\cal D}^b{\cal D}_{[b}\Theta(A)_{a]}-2{\cal D}_{[a}{\cal D}_{b]}\Theta(A)^b\nonumber\\
&-\ft32 \rmi\left(2\mathcal{D}_{a}X^{I}\Theta(X)_{I}-\bar{\Omega}^{I}\overset{\leftrightarrow}{\mathcal{D}}_{a}\Theta(\Omega)_{I}-2F^{I}\mathcal{D}_{a}\Theta(F)_{I}-\hc\right)\nonumber\\
&+\Theta(W)_{a A}D^A+\ft32\rmi\bar\Theta(\lambda)_A\gamma_*\gamma_{a}{}^{b}\overset{\leftrightarrow}{\mathcal{D}}_b\lambda^A\nonumber\\
&+3v_a\Theta(L)+3v^b\Theta(a)_{ba}+3\rmi{\cal D}^b L\tilde{\Theta}(a)_{ba}-\ft32\rmi\bar\Theta(\chi)\gamma_*\gamma_{a}{}^b\overset{\leftrightarrow}{\mathcal{D}}_b\chi\,.
\label{compETMgeneral}
\end{align}
We will further restrict ourselves to the components up to ${\cal B}_{ab}$ arguing that the higher components follow from the constraint that the multiplet is a generalization of a linear multiplet, i.e.
\begin{equation}
  \overline{{\cal D}}^{\dot\beta }\frac{1}{L}\overline{{\cal D}}^{\dot\alpha }{\cal E}_{\alpha \dot \alpha }=0\,.
 \label{constraintlinETML}
\end{equation}
In the Poincar\'{e} limit, $L$ is a constant, and this equation reduces to $\bar D^2 E_{\alpha \dot \alpha }=0$.
The constraint (\ref{constraintlinETML}) applies also to the gauge multiplet, replacing $\frac{1}{L}$ with $\frac{1}{D}$.

\subsection{Components in old minimal formulation with FI term}
The values of the covariantized field equations depend on the considered action. For the action in section \ref{ss:omFI} in WZ gauge we have
\begin{align}
  \Theta (A)_a =&  - 3\rmi X^0{\cal D}_a \bar X^{\bar 0}+3\rmi \bar X^{\bar 0}{\cal D}_a X^0-\ft32\rmi \overline{\Omega }{}^0P_L\gamma _a \Omega ^{\bar 0}\,,\nonumber\\
  \Theta (\psi )_a= & \ldots +\ft34\rmi\xi \gamma _a\left(X^0P_L-\bar X^{\bar 0}P_R\right)\lambda\,, \nonumber\\
  \Theta(e)_{ab}  =& \ldots +\ft32\eta _{ab}\xi \left[X^0\bar X^0 D+\sqrt{2}\rmi\bar \lambda \left(P_R\Omega ^{\bar 0}X^0-P_L\Omega ^0X^{\bar 0}\right)\right]\,, \nonumber\\
\Theta (F)_0=&-3\bar F^{\bar 0}\,,\nonumber\\
  \Theta (\Omega )_0=&3P_L\slashed{\cal D}\Omega^{\bar 0}-\ft32\sqrt{2}\rmi\xi \bar X^{\bar 0}P_L\lambda \,,\nonumber\\
\Theta (X)_0=& -3\bbox^c \bar{X}^{\bar 0}+\ft32\xi \left(-\bar X^{\bar 0}D+\sqrt{2}\rmi\bar \lambda P_R\Omega ^{\bar 0}\right)\,,\nonumber\\
  \Theta (D)=&-{\cal P}=-\ft32\xi X^0\bar X^{\bar 0}\,,\nonumber\\
  \Theta (\lambda )=& \ft32\sqrt{2}\rmi \xi \left(P_R\Omega ^{\bar 0}X^0-P_L\Omega ^0X^{\bar 0}\right)\,,\nonumber\\
\Theta (W)_a =& \ft32\rmi \xi\left(\bar X^{\bar 0}{\cal D}_aX^0-X^0{\cal D}_a\bar X^{\bar 0}+ \bar \Omega ^0\gamma _a P_R\Omega ^{\bar 0}\right)\,,
\label{ThetaomFI}
\end{align}
where the $\ldots $ are the expressions in (6.5) of \cite{Ferrara:2017yhz}, and the covariant derivatives also include the connection for (\ref{deltheta}) with gauge field $W_\mu $.

We consider the conformal gauge
\begin{equation}
 \mbox{Conformal gauge old minimal :}\quad\left.X^0\right|_\poinc=\kappa^{-1}\, , \qquad \left.\Omega^0\right|_\poinc=0\,, \qquad b_\mu =0\,,\label{gfconf}
\end{equation}
after which the conformal $\U(1)$ symmetries mix with the $\theta$ gauge symmetries, $\lambda _T= -\ft12\xi \theta $, and as such the fields transform with a different weight under the remaining gauge symmetries, as in (\ref{chargesfields}) in Appendix \ref{app:chiralsgcoupling}. Correspondingly, we define
\begin{equation}
  A^\xi _a = A_a +\ft12\xi W_a\,,
 \label{Aprime}
\end{equation}
which is thus inert under the remaining transformations, and also appears in the $S$-super\-symmetry decomposition law instead of $A_a $:
\begin{align}
  \delta _{\poinc}(\epsilon )=& \delta _Q(\epsilon )+ \delta _S\left(\eta = \ft12(\rmi\gamma _*\slashed{A}^\xi -P_R u-P_L\bar u)\epsilon \right)+\delta _{\rm K}\left(\lambda _{{\rm K}a}= -\ft14\bar \epsilon \hat{\phi }_a \right)\,,\nonumber\\
  &P_L\hat{\phi }_a=P_L\phi _a+\ft12 P_L(\rmi\slashed{A}^\xi +\bar u)\psi _a\,,\qquad u\equiv   \kappa \bar F^{\bar 0} \,.
 \label{delPoincare}
\end{align}

Then the field equations become
\begin{align}
  \kappa ^2\Theta(A)_a&\stackrel{\boxdot}{=}6A^\xi _a\,,\nonumber\\
\kappa ^2\Theta(\psi)_a &\stackrel{\boxdot}{=}-\ft12 \gamma_{abc}\widehat{R}^{bc}(Q)-\ft34 \rmi\xi \gamma _*\gamma _a\lambda \,,\nonumber\\
\kappa ^2 \Theta(e)_{ab}&\stackrel{\boxdot}{=}\widehat{G}_{ab}+6A^\xi _a A^\xi _b+3\eta_{ab}\left(u\bar{u}-A^\xi _cA^{\xi c}+\ft12D\right)\,,\nonumber\\
  \kappa \Theta (F^0)&=-3u\,,\nonumber\\
\sqrt{2}\kappa \Theta(\Omega^0) &\stackrel{\boxdot}{=} P_L\gamma ^{ab}\widehat{R}_{ab}(Q)-3\rmi\xi P_L\lambda \,,\nonumber\\
 \kappa  \Theta(X^0) &\stackrel{\boxdot}{=} -3\rmi\widehat{{\cal D}}^a A^\xi _a+\ft12 \widehat{R}+3A^{\xi  a} A^\xi _a-\ft32\xi D\,,\nonumber\\
\kappa ^2 \Theta (D)&\stackrel{\boxdot}{=}-\ft32\xi\,,\qquad \Theta (\lambda )\stackrel{\boxdot}{=}0\,,\qquad  \kappa ^2\Theta (W)_a\stackrel{\boxdot}{=}3\xi A^\xi _a\,,
\label{Thetaconfgauge}
\end{align}
where
\begin{align}
\widehat{R}_{\mu \nu  }(Q) =& 2\left( \partial _{[\mu }-\ft32\rmi
A_{[\mu }\gamma _*+\ft14\omega _{[\mu }{}^{ab}(e,\psi )\gamma _{ab}+\ft12 \gamma _{[\mu }(-\rmi\gamma _*\slashed{A}^\xi +\bar u)P_L+uP_R\right) \psi
_{\nu ]} \,, \nonumber\\
  \widehat{{\cal D}}_a A^\xi _b=&\nabla _aA^\xi _b+\ft12\rmi\bar \psi _a\gamma _* \hat{\phi }_b\,,\qquad \hat{\phi }_a= \ft12\gamma ^b\widehat{R}_{ba}(Q)+\ft1{12}\gamma _a\gamma ^{bc}\widehat{R}_{bc}(Q)\nonumber\\
\widehat{G}_{ab}=&\widehat{R}_{ab}-\ft12\eta  _{ab}\eta ^{cd}\widehat{R}_{cd}
\qquad \widehat{R}_{ab}=R_{(ab)}-\ft12\bar \psi^c \gamma _{(a} \widehat{R}_{b)c}(Q)+\ft12\bar \psi _{(a}\gamma ^c\widehat{R}_{b)c}(Q)\,,
\label{covRicci}
\end{align}
and $R_{ab}$ is the torsionful Ricci tensor.

The first components of the Einstein curvature multiplet (\ref{compETMgeneral}) in the gauge (\ref{gfconf}) are then
\begin{align}
\kappa^2{\cal C}_a=&6A^\xi _a\,,\nonumber\\
\kappa^2{\cal Z}_a=&-\ft3{2}\gamma^{\mu\nu\rho}\widehat{R}_{\nu\rho}(Q)-\gamma_a\gamma^{\mu\nu}\widehat{R}_{\mu\nu}(Q)+3\rmi\gamma_*\gamma_a\lambda\xi\,,\nonumber\\
\kappa^2{\cal H}_a=&-6\rmi\widehat{\cal D}_a \bar u\,,\qquad \widehat{\cal D}_a \bar u\equiv  (\partial _a-\ft32\rmi \xi W_a)\bar u+\bar \psi _aP_R\gamma \cdot \hat\phi \,,\nonumber\\
\kappa ^2{\cal B}_{ba}=&3\widehat{G}_{ab}-\eta_{ab}\widehat{G}_c{}^c+18A^\xi _aA^\xi _b+3\eta_{ab}\left(u\bar u- A^\xi _cA^{\xi c}-\xi D\right)\nonumber\\
& +3\varepsilon_{abcd}\widehat{\cal D}^c A^{\xi d}-\ft92\rmi\xi \widetilde{F}_{ab}(W)\,.
\label{compEom}
\end{align}

\subsection{Components in new minimal supergravity}
Since $A_\mu $ does not appear in the action (\ref{SFI}), the first component of the Einstein multiplet, proportional to $\Theta (A)_a$, does not get a contribution from the FI term, and the Einstein tensor multiplet is unchanged.\footnote{Other field equations, like $\Theta (\psi )$, $\Theta (e)$, do change but these contributions cancel since the first component of the multiplet is the same.}

We immediately consider the conformal gauge
\begin{equation}
 \mbox{Conformal gauge new minimal :}\quad\left.L\right|_\poinc=\kappa^{-2}\, , \qquad \left.\chi \right|_\poinc=0\,, \qquad \left.b_\mu\right|_\poinc =0\,,\label{gfconfnm}
\end{equation}
which implies that the Poincar\'{e} supersymmetry transformations are
\begin{align}
  \delta _{\poinc}(\epsilon )=& \delta _Q(\epsilon )+ \delta _S\left(\eta = \ft14\rmi\kappa^2 \gamma _*\slashed{v}\epsilon\right)+\delta _{\rm K}\left(\lambda _{{\rm K}a}= -\ft14\bar \epsilon \hat{\phi }_a \right)\,,\nonumber\\
  &\hat{\phi }_a=\phi _a-\ft14 \rmi\kappa ^2\gamma _*\slashed{v}\psi _a \,.
 \label{delPoincarenm}
\end{align}
and $\hat{\phi }_a$ is of the same form as in  (\ref{covRicci}), but with
\begin{align}
\widehat{R}_{\mu\nu}(Q)&=2\left(\partial_{[\mu}-\ft32\rmi A_{[\mu}\gamma_*+\ft14\omega_{[\mu}{}^{ab}\gamma_{ab}+\ft14\rmi\gamma_*\gamma_{[\mu}\slashed{H}\right)\psi_{\nu]}\,,\nonumber\\
H_a&= \kappa ^2v_a= e^{-1}e_{a\mu}\varepsilon^{\mu\nu\rho\sigma}\left(\kappa^2\partial_{\nu}a_{\rho\sigma}-\ft1{4}\bar\psi_\nu\gamma_\rho\psi_\sigma\right)\,.
\end{align}
The action (\ref{SL}) in this Poincar\'{e} form is
\begin{align}
\left[3L\ln\frac{L}{X^0\bar{X}^0}\right]_D\stackrel{\boxdot}{=}\int&\mathrm{d}^4 x\, \rme \frac1{2\kappa^2} \,\left[R-\ft1{2}\bar{\psi}_\mu\gamma^{\mu\nu\rho}\widehat{R}_{\nu\rho}(Q)+\ft{3}{2}H_aH^a\right]\nonumber\\
&+\ft32\varepsilon^{\mu\nu\rho\sigma}(2A_\mu- H_\mu)\partial_\nu a_{\rho\sigma}\,.\label{NM}
\end{align}
The action (\ref{SL}) leads to the following field equations
\begin{align}
\Theta(L)&\stackrel{\boxdot}{=}\ft12\eta ^{ab}\widehat{R}_{ab}+\ft34H_aH^a\,,\nonumber\\
\Theta(\chi)&=\frac3{2L}\slashed{\cal D}\chi+\frac32P_L\chi\bar\chi P_R\chi+¨\frac32P_R\chi\bar\chi P_L\chi-\frac{3}{4L^2}\rmi \gamma_*\slashed{v}\chi\nonumber\\
&\stackrel{\boxdot}{=}3\rmi\gamma^a\gamma_*\hat\phi_a=\frac12\rmi \gamma_*\gamma\cdot \widehat{R}(Q)\,,\nonumber\\
\kappa ^2\Theta(e)_{ab}&\stackrel{\boxdot}{=}\widehat{G}_{ab}-\frac34\eta_{ab}H_c H^c+\frac32 H_a H_b\,,\nonumber\\
\Theta(a)_{ab}&\stackrel{\boxdot}{=}3\rmi\tilde F_{ab}(A)-\frac32 \varepsilon_{abcd}\widehat{\cal D}^cH^d\,,\nonumber\\
\kappa^2\Theta(\psi)_a&\stackrel{\boxdot}{=}-\ft1{2}\gamma_{a}{}^{bc}\widehat{R}_{bc}(Q)\,,\nonumber\\
\Theta(A)_a&=-\frac9{8L}\rmi\bar\chi\gamma_*\gamma_a\chi+3v_a\stackrel{\boxdot}{=}\frac3{\kappa^{2}} H_a\,.
\end{align}
The first components of the Einstein curvature multiplet (\ref{compETMgeneral}) in the gauge (\ref{gfconfnm}) are then
\begin{align}
\kappa ^2{\cal C}_a&=3H_a\,,\nonumber\\
\kappa^2{\cal Z}_a&=-\ft{3}{2}\gamma_a{}^{bc}\widehat{R}_{bc}(Q)\,,\nonumber\\
{\cal H}_a&=0\,,\nonumber\\
\kappa ^2{\cal B}_{ba}&=3\left(\widehat{G}_{ab}-\ft34\eta_{ab}H_c H^c+\ft32H_a H_b-\varepsilon_{abcd}\widehat{\cal D}^cH^d+3\rmi\tilde F_{ab}(A)\right)\,.
\label{compEnm}
\end{align}

\section{Curved backgrounds}
\label{ss:apply}

To obtain supersymmetric backgrounds that preserve four supersymmetries, we consider the three curvature multiplets: Weyl, Einstein and scalar curvatures and impose that the high $\theta $ components (in presence of auxiliary fields) are vanishing in the background. For the Weyl multiplet, $W_{\alpha \beta \gamma }$, the vanishing of the second component leads to the vanishing of the Weyl tensor
\begin{equation}
  W_{\mu \nu \rho \sigma }=0\,,
 \label{Weyl0}
\end{equation}
which will be valid for all the examples below.

Since the scalar multiplet follows from the transformation of the Einstein multiplet, the other constraints follow already from the latter.
In particular we have to consider the transformation of the fermion field ${\cal Z}_a$ in the Einstein multiplet:
\begin{equation}
  P_L \delta {\cal Z}_a= \ft12 \kappa ^2P_L\left( \rmi{\cal H}_a -\gamma ^b{\cal B}_{ba}-\rmi \slashed{\cal D}{\cal C}_a\right)\epsilon +\rmi P_L\left(-3{\cal C}_a +\gamma _{ab}{\cal C}^b\right)\eta \,.
 \label{PLdelzetaa}
\end{equation}
We can write the $S$ supersymmetry parameter of  (\ref{delPoincare}) and  (\ref{delPoincarenm}) in a uniform way using the values of ${\cal C}_a$ in (\ref{compEom}) and (\ref{compEnm}):
\begin{equation}
 P_L\eta = \left(\ft1{12}\rmi\kappa ^2\slashed{\cal C} -\ft12\bar u\right)\epsilon\,,
 \label{etauniform}
\end{equation}
just taking $u=0$ in case of new minimal. This leads to
\begin{align}
 P_L \delta {\cal Z}_a= &
 \ft12 P_L\left[ \rmi({\cal H}_a+3\kappa ^2{\cal C}_a\bar u) -\gamma ^b\kappa ^{-2}B_{ba}-\rmi \slashed{\cal D}{\cal C}_a-\rmi\gamma _{ab}{\cal C}^b\bar u\right]\epsilon\,,
\label{PLdelazetaPoinc}
\end{align}
where
\begin{equation}
B_{ba}\equiv \kappa ^2{\cal B}_{ba}+\ft16\kappa ^4\left(-4{\cal C}_a{\cal C}_b+\eta _{ab}{\cal C}_c{\cal C}^c\right)\,.
 \label{BPoincdefined}
\end{equation}
The vanishing of the right-hand side of (\ref{PLdelazetaPoinc}) constrains supersymmetric backgrounds. Splitting in independent parts we get
\begin{align}
 {\cal C}_a u=0\,,\qquad {\cal H}_a=0\,,\qquad B_{ba}=0\,,\qquad {\cal D}_a{\cal C}_b=0\,.
 \label{generalSUSYbg}
\end{align}
We can then further consider only the bosonic part of these equations.

In the previous section, we did not include here the contributions of the gauge kinetic terms, see section \ref{ss:gaugekin}. Note, however, that (\ref{EWa}) implies that ${\cal C}_a$ does not have bosonic terms from this part, and from (\ref{compETMgeneral}) we see that also ${\cal H}_a$ does not get extra bosonic terms. There will be new terms in $\Theta (e)_{ab}$, whose bosonic part is
\begin{equation}
  \Theta^W (e)_{ba}= F_{c(a}(W)F_{b)}{}^c(W) + \ft14\eta _{ab}F_{cd}(W)F{}^{cd}(W)-\ft12\eta _{ab}D^2\,.
 \label{Thetaegauge}
\end{equation}
Taking also into account (\ref{ThetaWD}), this gives from (\ref{compETMgeneral}) an extra contribution to ${\cal B}_{ba}$:
\begin{equation}
 {\cal B}_{ba}^W = 3F_{c(a}(W)F_{b)}{}^c(W) + \ft34\eta _{ab}F_{cd}(W)F{}^{cd}(W)+\ft32\eta _{ab}D^2+3\rmi\tilde F_{ab}(W)\,D\,.
 \label{BfromW}
\end{equation}

\subsection{Curved backgrounds old minimal}
The equations in (\ref{generalSUSYbg}), apart from $B_{ba}=0$, imply with the bosonic parts of the values in (\ref{compEom}):
\begin{equation}
  u A^{\xi}_\mu =0\,,\qquad \left( \partial _\mu +\ft32\rmi\xi  W_\mu\right)u=0\,,\qquad \nabla _\mu A^\xi _\nu =0\,.
 \label{firsteqnsom}
\end{equation}
Then the antisymmetric part of $B_{ba}=0$, has only the terms with $\tilde F_{ab}(W)$, which vanish adding (\ref{compEom}) to (\ref{BfromW}) upon use of the $D$ field equation
\begin{equation}
  D\approx {\cal P}=  \ft32\xi X^0 \bar X^0 \stackrel{\boxdot}{=}\ft32\xi\kappa ^{-2} \,.
 \label{Deom}
\end{equation}

In this case, we get from (\ref{BPoincdefined}), without taking into account the gauge terms (\ref{BfromW}), that the symmetric part of $B_{ab}$ is
\begin{align}
B_{(ba)}=3\widehat{G}_{ab}-\eta_{ab}\widehat{G}_c{}^c-6A^\xi _aA^\xi _b+3\eta_{ab}\left(u\bar u+ A^\xi _cA^{\xi c}-\xi D\right)\,.
\end{align}
Hence, without matter, the vanishing of this expression leads to
\begin{equation}
   R_{ab}-2A^\xi_a A^\xi_b +\eta _{ab}(2A^{\xi c} A^\xi_c +3u\,\bar u-3\xi D)=0\,.
 \label{Rmunueqnom}
\end{equation}
The above equations for $\xi =0$ were also obtained in section 6.1 of \cite{Ferrara:2017yhz} and agree with  \cite{Festuccia:2011ws,Dumitrescu:2012ha,Dumitrescu:2012at}.
Note that $R+ 6A_a A^a + 12 u\,\bar u=0$ and $\nabla ^\mu A_\mu =0$ are the vanishing of the upper component of the multiplet ${\cal R}$.
The equations allow supersymmetric solutions\footnote{In \cite{Festuccia:2011ws,Dumitrescu:2012ha,Dumitrescu:2012at} also Euclidean theories are considered by treating real supergravity fields as complex fields. Such an extension is not included in the investigation in this paper.}
\begin{align}
  \mbox{AdS}_4:\ & A_a \approx 0\,,\qquad R+12u\,\bar u \approx 0\,,\qquad u\neq 0\,,\nonumber\\
S^3\times L:\ & A_a\approx (A_0, \, 0, \, 0, \, 0)\,, \qquad u\approx 0\,, \qquad B_{ab}\approx0\,,\nonumber\\
&R_{00}\approx R_{0i}\approx 0, \quad R_{ij}\approx \phantom{-}2 A_0^2\delta_{ij}\,, \qquad (i=1,2,3)\,,\qquad R\approx\ 6A_0^2\,,\nonumber\\
AdS_3\times L:\ & A_a\approx (0, \, 0, \, 0, \, A_3)\,, \qquad u\approx 0\,, \qquad B_{ab}\approx 0\,,\nonumber\\
&R_{33}\approx R_{3i}\approx 0, \quad R_{ij}\approx -2 A_3^2\eta _{ij}\,, \qquad (i=0,1,2)\,,\qquad R\approx -6A_3^2\,,
\label{solnoldmin}
\end{align}
where $A_0$ or $A_3$  are constants.

For a de Sitter configuration, an extra nilpotent Volkov--Akulov field $X^1$ was introduced in \cite{Ferrara:2017yhz} and a possible solution is
\begin{align}
  \mbox{ dS: } & F^0 \approx   \kappa^{-2} \lambda\,, \qquad F^1 \approx -\ft13 \kappa^{-2}\mu\, , \nonumber\\
& \kappa {\cal R}|_{\text{last}}=-\ft16 R \kappa -2 \kappa^3 |F_0|^2 \approx
 -\ft 29 \kappa^{-1} \mu^2\,, \qquad  \left.\kappa{\cal R}\right\vert_{\text{first}} \approx \lambda\,.
\label{dSnewmin}
\end{align}
The scalar curvature is then $R\approx 4\kappa ^2 V= \ft43\kappa ^{-2}\mu ^2-12\kappa ^{-2}\lambda ^2$  and thus
\begin{equation}
  B_a{}^a \approx R +12 u\bar u\approx 
  16\kappa ^{-2}\mu ^2\,,
 \label{BtracedS}
\end{equation}
signals the breaking of supersymmetry.

\subsection{FI term in components}
\label{ss:FIcomponents}
In absence of matter ($S^i$) the FI model is the Freedman model \cite{Freedman:1976uk,deWit:1978ww}, which has a field equation (\ref{ThetaV})
\begin{equation}
  T(\rme^{\xi V}\bar X^0)\approx 0\,.
 \label{feF0V}
\end{equation}
This is the covariantized ${\cal R}$ chiral scalar curvature multiplet. Note that the above equation in components and in the WZ gauge reproduces the field equations. The first component is
\begin{equation}
 u=\kappa \bar F^0\approx 0\,,
 \label{firstcovR}
\end{equation}
while the  last component is proportional to $\Theta(X)_0$ in  (\ref{ThetaomFI})
\begin{equation}
  \left(\partial ^\mu +\rmi A^{\xi \mu}  \right)\left(\partial _\mu +\rmi A^\xi _\mu  \right)\bar X^0-\ft16 R\bar X^0+\ft12\xi\bar X^0\,D\approx \mbox{fermionic terms}
 \label{lastcovR}
\end{equation}
With zero fermions and in conformal gauge a de Sitter solution is obtained with
\begin{equation}
  A^\xi _\mu  \stackrel{\boxdot}{\approx }0\,,\qquad
  R\approx 3\xi D \stackrel{\boxdot}{\approx } \ft92\kappa ^{-2}\xi ^2\,.
 \label{RxiD}
\end{equation}
The breaking of supersymmetry is evident from
\begin{equation}
  B_a{}^a \approx R - 12\xi D \approx -9\xi D\,.
 \label{BtracedSxi}
\end{equation}

\subsection{Curved backgrounds new minimal}
\label{ss:curvedbgnm}
In this case, with the values in (\ref{compEnm}), the first two equations of (\ref{generalSUSYbg}) are empty. The last equation and the antisymmetric part of $B_{ab}$ imply
\begin{equation}
 \nabla_{\mu} H_{\nu}=0\,,\qquad \partial_{[\mu} A_{\nu]}=  0\,,
 \label{susybgnm}
\end{equation}
The symmetric part of $B_{ab}$ is in this case
\begin{align}
B_{(ba)}=3\left(\widehat{G}_{ab}-\ft12H_a H_b-\ft14\eta_{ab}H_c H^c\right)\,.
\end{align}
and its vanishing implies
\begin{align}
R_{ab}=&\ft12\left(H_aH_b-\eta_{ab}H^cH_c\right)\,.
\label{FSnm}
\end{align}
The trace of the latter, $R+\ft32H^aH_a=0$ is also the $\theta $ component of the chiral curvature.
These equations agree with the constraints in \cite{Festuccia:2011ws}. 
The solutions $S^3\times L$ and $AdS_3\times L$ in (\ref{solnoldmin}) are also possible here, in this case with $A_\mu $ vanishing, and $H_a/2$ taking the value of $A_a$ of the old-minimal solutions. However, there is no room for the $AdS_4$ solution in this case.
These are just examples of solutions of the supersymmetry-preservation conditions.

\section{Auxiliary fields and conformal compensators}
\label{app:auxf}
We give here a summary and discussion of auxiliary field formulations of ${\cal N}=1$, $D=4$ supergravity. First of all we repeat the argument that auxiliary field formulations that have a consistent Lagrangian and describe pure supergravity can only have $(12+12)$ + a multiple of $(8+8)$ degrees of freedom. This is so because additional fermions should come in pairs of 4-component spinors in the form ${\cal L}= \ldots +\bar \lambda \chi +\ldots $. This is e.g. also mentioned in  \cite{Sohnius:1982xs}, where they clearly mention that their $(16+16)$ set has either to be reduced to one of the $(12+12)$ sets, or it describes additional physical degrees of freedom. Similarly, the structure in \cite{deWit:1978ww} with $(16+16)$ fields is in fact an off-shell description of the Freedman model \cite{Freedman:1976uk}, discussed also in this paper, and thus contains other propagating degrees of freedom.

Still there have been made claims of sets of auxiliary field different from the old minimal \cite{Ferrara:1978em,Stelle:1978ye,Fradkin:1978jq}, new minimal \cite{Sohnius:1981tp}, or non-minimal set \cite{Breitenlohner:1976nv,Breitenlohner:1977jn}. There have been proposals for new non-minimal sets of auxiliary fields or $(16+16)$ sets. The latter suffer certainly from the feature mentioned above that they cannot describe pure supergravity. Still one may wonder whether the theory with extra physical degrees of freedom can be obtained using the minimal sets of auxiliary fields.

The known sets of auxiliary fields can be obtained by coupling the Weyl multiplet to a compensating multiplet, and perform gauge fixings.
The old minimal supergravity is obtained by using a chiral compensating multiplet \cite{Kaku:1978ea,Ferrara:1978rk}, new minimal supergravity is obtained by a real linear compensator \cite{Siegel:1978mj,deWit:1981fh} and non-minimal supergravity by a complex linear multiplet \cite{Siegel:1978mj,deWit:1981fh}. The latter has an arbitrary Weyl weight $w$ and for generic Weyl weight the multiplet is irreducible. For Weyl weight $w=2$ the real part can be separated, and the multiplet can be restricted to a real linear multiplet, and for $w=0$, a chiral multiplet can be separated (by acting with $T$ on the complex linear multiplet). For these values of the parameter the non-minimal set thus reduces to one of the minimal sets plus a meaningless separate part.

Theories with the same set of multiplets but different gauge fixings are equivalent, i.e. different gauge fixings can always be rephrased as field redefinitions \cite{deWit:1982na}. In \cite{Kugo:1983ym} it has been checked that the new non-minimal sets are reproduced by a gauge fixing of a reducible set of compensating multiplets, containing a chiral multiplet or a linear multiplet, and another multiplet. Hence, by
choosing a gauge fixing in the chiral or linear multiplet, these set of auxiliary fields become reducible, and one can restrict to one of the minimal sets of auxiliary fields. Later the $(16+16)$ sets were brought up \cite{Girardi:1984vq,Lang:1985xk}. However, it was shown again that these are just obtained by a different gauge fixing when using together a chiral and a real linear multiplet \cite{Siegel:1986sv,Hayashi:1985vd,Aulakh:1985dn}. These authors reproduced the action and transformation law of the $(16+16)$ set by some gauge choice, and showed by another gauge choice that the theory can be reduced to one of the minimal versions coupled to a physical real linear or chiral multiplet. They even exhibited explicitly the redefinitions that are equivalent to these changes of gauge conditions.

This shows that the known ${\cal N}=1$ sets of auxiliary fields can be restricted to the old and new minimal and the non-minimal ones, each described by one irreducible compensating multiplet.
In this paper we considered old and new minimal sets. Then the auxiliary fields are
\begin{align}
  \mbox{old minimal: } & (A_a ,\, u=\kappa \bar F^0)\,, \nonumber\\
  \mbox{new minimal: } & (A_\mu ,\, a_{\mu \nu })\,,\qquad H^\mu \equiv  e^{-1}\varepsilon^{\mu\nu\rho\sigma}\left(\kappa^2\partial_{\nu}a_{\rho\sigma}-\ft1{4}\bar\psi_\nu\gamma_\rho\psi_\sigma\right)\,.
\label{auxfields}
\end{align}
The actions for pure Poincar\'{e} supergravity are
\begin{align}
 \kappa ^2 e^{-1} {\cal L}^{\rm om} =& \frac12 R-\frac12\bar \psi ^\mu  \gamma ^{\mu \nu \rho }D_\nu \psi _\rho -\frac13 u\,\bar u + 3A^aA_a \nonumber\\
 \kappa ^2 e^{-1}  {\cal L}^{\rm nm}  = & \frac12 R-\frac12\bar \psi ^\mu  \gamma ^{\mu \nu \rho }(D_\nu-\ft32\rmi\gamma _*A_\nu ) \psi _\rho +\frac34H_aH^a+\frac32\kappa ^2\varepsilon^{\mu\nu\rho\sigma}(2A_\mu- H_\mu)\partial_\nu a_{\rho\sigma}\,.
\label{Lomnm}
\end{align}

\section{Conclusion}
\label{ss:conclusion}
We obtained the conservation laws in old and new minimal supergravity, also in case that there are FI terms. They can be written as equations on conformal superprimary multiplets. They follow from Ward identities that can be derived in general. Such general relations allow us also to obtain the different components of the Einstein supermultiplet. This in turn gives an alternative derivation of the Festuccia--Seiberg relations  \cite{Festuccia:2011ws} for supersymmetric backgrounds.

One of the main results of the present paper is the Ward Identity given by (\ref{DEoldminFIVintro}), which shows that  (\ref{conservationEqn}) is perfectly compatible with the old minimal formulation provided one realises that the FI gauging makes the chiral compensator not inert and the mere existence of the FI gauge field  introduces eight more (4B+4F) off-shell degrees of freedom.
This is the reason why the authors of  \cite{deWit:1978ww} called it chirally extended Supergravity, although it was originally introduced
in the context of the old minimal formulation by Stelle and West \cite{Stelle:1978wj}.
We point out that these results are the local exact nonlinear version of the linearized results of  \cite{Komargodski:2010rb} obtained by coupling the
rigid $S$ super current multiplet to linearised supergravity. A similar modification is expected to work also for the K\"{a}hler invariance. The modification of the super current conservation law occurs since the chiral compensator is not inert under FI or K\"{a}hler symmetry. Such modification doesn't occur in new minimal supergravity because the linear multiplet
compensator is inert under these gauge symmetries (see (\ref{fechitotal})).

\bigskip
\section*{Acknowledgments}

\noindent We would like to thank F. Farakos for discussions.
The work of S.F. is supported in part by CERN TH Dept and INFN-CSN4-GSS. M.S. is grateful to the CERN TH Dept and the Physics Department of Turin University for hospitality during the initial stages of this work.
The work of M.T. and A.V.P. is supported in part by the KU Leuven C1 grant ZKD1118 C16/16/005. The work of M.T. is supported by the FWO odysseus grant G.0.E52.14N.

\appendix
\section{Linear multiplet and related spinor chiral superfield}
\label{app:linWL}
For convenience, we repeat the transformations of the independent fields of the linear multiplet
\begin{align}
\delta L &=\frac12\rmi\bar\epsilon\gamma_*\chi\,,\nonumber\\
\delta \chi &=-\frac12\left(\slashed{v}+\rmi\gamma_*\slashed{\cal D}L\right)\epsilon-2\rmi\gamma _*\eta\,L \,,\nonumber\\
\delta a_{\mu\nu}&=\frac14\rmi\bar\epsilon\gamma_{\mu\nu}\gamma_*\chi+\frac12\bar\epsilon\gamma_{[\mu}\psi_{\nu]}L\,.
\end{align}
where
\begin{equation}
  v_a=\rme^{-1} e_{a\mu}\varepsilon^{\mu\nu\rho\sigma}\left(\partial_\nu a_{\rho\sigma}-\frac14\bar{\psi}_\nu\gamma_\rho\psi_\sigma L\right)+\frac12\bar\chi\gamma_{ab}\psi^b\,.
 \label{defva}
\end{equation}

Therefore, (\ref{delThetacov}) implies that in any action containing this multiplet
\begin{equation}
  \delta (\epsilon )P_L\Theta (\chi )= -\ft12\rmi P_L\epsilon \Theta (L) + \ft14 \rmi \gamma ^{\mu \nu }P_L\epsilon \Theta (a)_{\mu \nu }\,.
 \label{delepsThetachi}
\end{equation}
First of all this implies that $P_L\Theta (\chi )$ is a chiral multiplet. It does not transform under S-supersymmetry, and is thus a super-primary field. Since $P_L\chi $ has (Weyl,chiral) weights $(5/2,-3/2)$, the field equation $P_L\Theta (\chi) $  has weights $(3/2,3/2)$, consistent with the requirements for a chiral multiplet \cite{Kugo:1983mv,Ferrara:2016een}.
Furthermore we find
\begin{equation}
  {\cal D}^\alpha \Theta (\chi )_\alpha =\rmi\Theta (L)\,.
 \label{DThetachi}
\end{equation}
Therefore it is convenient to define
\begin{equation}
  W_\alpha^{\chi } \equiv  \rmi \Theta (\chi )_\alpha\,,
 \label{defWL}
\end{equation}
such that $D^\alpha W_\alpha^\chi = -\Theta (L)$ is real. The components of this chiral multiplet are
\begin{equation}
W_\alpha^{\chi }=\left\{\rmi\Theta(\chi)_\alpha ,\,\frac1{\sqrt2} (P_L)_{\beta\alpha}\Theta(L)+\frac1{2\sqrt2}(\gamma^{ab})_{\beta\alpha}\Theta(a)_{ab},\,-\rmi(\slashed{\cal D}\Theta(\chi))_\alpha\right\}\,.
\label{Wmultiplet}
\end{equation}

\section{Superconformal Ward identities}
\label{ss:WardId}
To derive the components of the Einstein curvature multiplet, we made use of the Ward identities that are granted by the superconformal symmetries. In our formalism Ward identities are constructed from the action in the following way\footnote{After partial integrating terms with $\partial _\mu \epsilon ^A$, extra curvature terms can appear in (\ref{WTA}), from (D.12) in \cite{Ferrara:2017yhz}.}
\begin{equation}
W(T^A)=\left(\delta(\epsilon^A)\phi^i\right)\frac{\delta S}{\delta \phi ^i}=\left(\delta(\epsilon^A)\phi^i\right)\Theta(\phi)_i+ \mbox{curvature terms}=0\,.
\label{WTA}
\end{equation}
The $\phi^i$ are the independent fields of the action and $T^A$ the operator of one of the symmetries. The Ward identities transform into Ward identities under the symmetries and are therefore invariant. We will now summarize the Ward identities for the conformal supergravity algebra
\begin{align}
	\text{Cov. gct: } W(P)_a\equiv&\;{\cal D}^{b}\Theta(e)_{ab}+\Theta(A)^{b}R_{ab}(T)+\ft34\bar{\Theta}(\psi)^{b}R_{ab}(Q)\nonumber\\
	&\;+\left[\Theta(X)^I {\cal D}_a X_I+\bar{\Theta}(\Omega)^I {\cal D}_a \Omega_I+\Theta(F)^I {\cal D}_a F_I+\hc\right]\nonumber\\
	&\;+\Theta(W)_{A}^b F_{ab}{}^A+\bar\Theta(\lambda)^A{\cal D}_a\lambda_a+\Theta(D)^A{\cal D}_a D_A\nonumber\\
	&\;+\Theta(L){\cal D}_a L+\bar\Theta(\chi){\cal D}_a\chi+3\Theta(a)^{bc}{\cal D}_{[a} a_{bc]} \,, \label{Ward P-sym}\\
	\text{Lorentz: } W(M)_{ba}\equiv&\;\Theta(e)_{[ba]}+\ft14\left[\bar{\Omega}^I\gamma_{ba}\Theta(\Omega)_I+\hc\right] +\ft14\bar\lambda^A\gamma_{ba}\Theta(\lambda)_A +\ft14\bar\chi\gamma_{ba}\Theta(\chi)\,, \label{Ward M-sym}\\
	\text{Dilatations: } W(D)\equiv&\;\Theta(e){_a}{}^{a}\nonumber\\ &+\left[w_I X^I\Theta(X)_I+(w_I+\ft12)\Omega^I \Theta (\Omega)_I+(w_I+1)\Theta (F)^I F_I
	+\hc\right]\nonumber\\ &+\ft32\bar\Theta(\lambda)^A\lambda_A+2\Theta(D)^A D_A +2L\Theta(L)+\ft52\bar\chi\Theta(\chi)\,,\label{Ward D-sym}\\
	\text{spec.conf.: } W(K)_a\equiv&\;\Theta (b)_a\,, \label{Ward K-sym}\\
	\text{$T$-symmetry: } W(T)\equiv&-{\cal D}^a\Theta (A)_a+\rmi\left[X^I\Theta (X)_I -\ft12\overline{\Omega} ^I\Theta (\Omega )_I - 2F^I\Theta (F)_I -\hc\right]\nonumber\\
	&\;+\ft32\rmi\bar\Theta(\lambda)^A\gamma_*\lambda_A-\ft32\rmi\bar\chi\gamma_*\Theta(\chi)\,, \label{Ward T-sym} \\
	\text{$Q$-susy: } W(Q)\equiv&\;-{\cal D}^a\Theta(\psi)_a \\
	&\;+ \ft{1}{\sqrt{2}}\left[\Omega^I\Theta(X)_I+\left(-\slashed{\cal D}X^I+F^I\right)\Theta(\Omega)_I+\slashed{\cal D}\Omega^I \Theta(F)_I+\hc\right]\nonumber\\
	&\; -\ft12\gamma_a\lambda^A\Theta(W)^a_{A}+\ft12\left(-\ft12\gamma\cdot F^A+\rmi\gamma_* D^A\right)\Theta(\lambda)_A\nonumber\\
	&\;+\ft12\rmi\gamma_*\slashed{\cal D}\lambda^A\Theta(D)_A+\ft12\rmi\gamma_*\chi\Theta(L)+\ft12\left(\slashed{v}-\rmi\gamma_*\slashed{\cal D}L\right)\Theta(\chi)\nonumber\\
	&\;+\ft12\left(\ft12\rmi\gamma_{\mu\nu}\gamma_*\chi+\gamma_{[\mu}\psi_{\nu]}L\right)\Theta(a)^{\mu\nu}\,, \label{Ward Q-sym}\\
  \text{$S$-susy: } W(S)\equiv&\;\gamma _a\Theta (\psi )^a + \sqrt{2}\left[X^I\Theta (\Omega )_I+\hc\right]-2\rmi\gamma_*\Theta(\chi)L\,.  \label{Ward S-sym}
\end{align}

\section{Current conservation with gauge symmetry: electrodynamics analogue}
\label{ss:currentem}
In this section, we understand the presence of the $W_\alpha $ contribution in (\ref{DJmatterFI}) in the context of electrodynamics. After all this contribution is not so surprising because in electrodynamics the conservation of the gauge-invariant matter stress tensor gets a contribution
\begin{equation}
  \partial ^\mu T_{\mu \nu }^{\rm{M}} \approx - F_{\nu \rho }J^{\rho \,\rm{M}}\,,
 \label{EManalogue}
\end{equation}
which just cancels the $\partial ^\mu T_{\mu \nu }^{\rm{e.m.}}$, which satisfies
\begin{equation}
   \partial ^\mu T_{\mu \nu }^{\rm{e.m.}} =2 F_{\nu \rho }\partial ^\lambda F_\lambda {}^{\rho}\,.
 \label{dTem}
\end{equation}
This corresponds to higher $\theta $ terms in (\ref{DJmatterFI}). The term $N^{\rm M}W_\alpha \propto \Phi ^{\rm M}W_\alpha $ contains at the $\theta^3 $ level the right-hand side of (\ref{EManalogue}) since $F_{\nu \rho }$ is at the $\theta $ level of $W_\alpha $ and at the $\theta ^2$ level of $\Phi ^{\rm M}$ is the component
\begin{equation}
  B_\mu (\Phi )= \rmi\left(\frac{\partial \Phi }{\partial S^i}\partial _\mu S^i - \frac{\partial \Phi }{\partial \bar S^i}\partial _\mu \bar S^i\right)=J_\mu ^{\rm M}\,.
 \label{Jmuattheta2}
\end{equation}

\section[Coupling chiral supergravity to a charged chiral multiplet]{Coupling chiral supergravity to a charged \\ chiral multiplet}
\label{app:chiralsgcoupling}

In this Appendix we give the Lagrangian when the supergravity with FI term is coupled to a charged chiral multiplet. The charge under the $\U(1)$ of the gauge multiplet is fixed by the requirement that a superpotential term be invariant. This analysis extends to an arbitrary number of charged fields \cite{Cremmer:1982en} but here we will confine the example to a single chiral multiplet $S$ with a superpotential
\newcommand{\paraml}{\ell}
\begin{equation}
  (X^0)^3 W(S)= \paraml (X^0)^3S\,,
 \label{WSlinear}
\end{equation}
linear in $S$, so that if $S^2=0$ is imposed we have the Volkov--Akulov (VA) theory.
The gauge-invariant supergravity action that includes (\ref{WSlinear}) is
\begin{equation}
  {\cal L}= \left[-3X^0\bar X^0\rme^{\xi V}\right]_D + \left[X^0\bar X^0S\bar S\rme^{-2\xi V}\right]_D +\paraml \left[(X^0)^3S\right]_F +\left[W^\alpha W_\alpha \right]_F\,,\qquad S^2=0\,,
 \label{FIVA}
\end{equation}
and the (FI) gauge transformations  (\ref{LambdaWZ}) contain then also a transformation for $S$:
\begin{equation}
  X^0\ \rightarrow\ X^0 \rme^{-\xi \Lambda }\,,\qquad S\ \rightarrow\ S \rme^{3\xi \Lambda }\,,\qquad V\ \rightarrow \ V + \Lambda +\bar \Lambda \,.
 \label{LambdaWZVA}
\end{equation}
In WZ gauge, this leads to transformations as in (\ref{deltheta}). Furthermore there are the conformal $\U(1)$ symmetries, e.g. $\delta X^0 = \rmi \lambda _T X^0$.
After the conformal gauge fixing that makes $X$ real, these transformations mix such that $X^0$ does not transform, i.e. $\lambda _T= -\ft12\xi \theta $. The fields that are charged under these symmetries, denoting the components of the multiplet $S$ as $\{S,\,P_L\chi ,\,f\}$, are
\begin{equation}
\begin{array}{|l|rr|r|}\hline
  \mbox{field} & \rmi\lambda _T & \rmi\xi \theta  & \mbox{resulting }\rmi\xi \theta  \\ \hline
\rule{0pt}{1\normalbaselineskip}  P_L\psi _\mu  & \ft32 & 0 & -\ft34  \\[1mm]
  P_L\lambda  & \ft32 & 0 & -\ft34  \\[1mm]
  X^0 & 1 & \ft12 & 0 \\[1mm]
  P_L\Omega^0  & -\ft12 & \ft12 & \ft34 \\[1mm]
  F^0 & -2 & \ft12 & \ft32 \\[1mm]
  S & 0 & -\ft32 & -\ft32 \\[1mm]
  P_L\chi  & -\ft32 & -\ft32 & -\ft34\\[1mm]
  f & -3 & -\ft32 & 0\\[1mm]
  D & 0 & 0 & 0\\[1mm] \hline
\end{array}
 \label{chargesfields}
\end{equation}
and opposite for the complex or charged conjugates.

The spectrum of this theory contains gravity coupled to Maxwell in de Sitter space with a charged scalar, a spin 1/2 field and a spin 3/2 massive gravitino.
The scalar disappears if $S$ is nilpotent.  This can be obtained by adding a chiral term $[\sigma S^2 (X^0)^3]_F$, with $\sigma $ a chiral multiplet with zero Weyl weight and transforming under (\ref{LambdaWZVA}) as $\sigma\rightarrow \sigma \exp(-3\xi \Lambda)$.
Thus the FI model can be written in a Volkov--Akulov form, as suggested in a final comment in \cite{deWit:1978ww}.
Deleting $\xi $ we have the Volkov--Akulov model coupled to supergravity \cite{Antoniadis:2014oya,Bergshoeff:2015tra}.

The Goldstino is then a linear combination of the gaugino $\lambda $ and the VA fermion $\chi $, since both $F$ and $D$ have a non-zero value. The cosmological constant is
\begin{equation}
 \kappa ^4 V =\ft98 \xi ^2 + |\paraml |^2\,,
 \label{VFIVA}
\end{equation}
with an additional charged scalar contribution if $S$ is not nilpotent.
By deleting $S$ we have the Freedman model, while by deleting $V$ (or setting $\xi =0$) we have pure supergravity  \cite{Ferrara:1978em} or Maxwell-Einstein supergravity with auxiliary fields  \cite{Cremmer:1982en}.

%
\providecommand{\href}[2]{#2}\begingroup\raggedright\endgroup

\end{document}